\documentclass[aps,preprint,prl,superscriptaddress,showpacs,amsmath,amssymb]{revtex4}

\usepackage{graphicx}
\usepackage{dcolumn}
\usepackage{bm}

\begin{document}

\title{Search for a stochastic background of 100-MHz gravitational waves with laser interferometers}

\author{Tomotada~Akutsu}
\email{takutsu@gravity.mtk.nao.ac.jp}
\affiliation{Department of Astronomy, University of Tokyo, Bunkyo, Tokyo 113-0033, Japan}
\author{Seiji~Kawamura}
\affiliation{National Astronomical Observatory of Japan, Mitaka, Tokyo 181-8588, Japan}
\author{Atsushi~Nishizawa}
\affiliation{Graduate School of Human and Environmental Studies, Kyoto University, Kyoto 606-8501, Japan}
\author{Koji~Arai}
\affiliation{National Astronomical Observatory of Japan, Mitaka, Tokyo 181-8588, Japan}
\author{Kazuhiro~Yamamoto}
\affiliation{National Astronomical Observatory of Japan, Mitaka, Tokyo 181-8588, Japan}
\author{Daisuke~Tatsumi}
\affiliation{National Astronomical Observatory of Japan, Mitaka, Tokyo 181-8588, Japan}
\author{Shigeo~Nagano}
\affiliation{National Institute of Information and Communications Technology, Koganei, Tokyo 184-8795, Japan}
\author{Erina~Nishida}
\affiliation{Graduate School of Humanities and Sciences, Ochanomizu University, Bunkyo, Tokyo 112-8610, Japan}
\author{Takeshi~Chiba}
\affiliation{Department of Physics, College of Humanities and Sciences, Nihon University, Tokyo 156-8550, Japan}
\author{Ryuichi~Takahashi}
\affiliation{Graduate School of Science, Nagoya University, Nagoya 464-8602, Japan}
\author{Naoshi~Sugiyama}
\affiliation{Graduate School of Science, Nagoya University, Nagoya 464-8602, Japan}
\author{Mitsuhiro~Fukushima}
\affiliation{National Astronomical Observatory of Japan, Mitaka, Tokyo 181-8588, Japan}
\author{Toshitaka~Yamazaki}
\affiliation{National Astronomical Observatory of Japan, Mitaka, Tokyo 181-8588, Japan}
\author{Masa-Katsu~Fujimoto}
\affiliation{National Astronomical Observatory of Japan, Mitaka, Tokyo 181-8588, Japan}

\date{\today}

\begin{abstract}
This letter reports the results of a search for a stochastic background of gravitational waves (GW) at 100~MHz by laser interferometry.
We have developed a GW detector, which is a pair of 75-cm baseline synchronous recycling (resonant recycling) interferometers.
Each interferometer has a strain sensitivity of $\sim 10^{-16}\,\mathrm{Hz^{-1/2}}$ at 100~MHz.
By cross-correlating the outputs of the two interferometers within $1000$ seconds, 
we found $h_{100}^2\Omega_{\mathrm{gw}}<6\times 10^{25}$ to be an upper limit
on the energy density spectrum of the GW background in a 2-kHz bandwidth around 100~MHz,
where a flat spectrum is assumed.
\end{abstract}

\pacs{Valid PACS appear here}
\maketitle

Recently, Cruise and Ingley reported on a detector for gravitational waves (GW) at 100 MHz~\cite{Cruise1}.
Their GW detector is a pair of waveguide loop cavities,
each of which has a strain sensitivity of $\sim 10^{-14}\,\mathrm{Hz^{-1/2}}$ at the frequency.
Except for this, no experiments were attempted to \textit{directly} detect GWs at very high frequencies (above 100 kHz),
while many theories predict a stochastic gravitational-wave background (GWB) in a broad range of frequencies, $10^{-18}-10^{10}\, \mathrm{Hz}$.
At very high frequencies, a relatively large GWB is predicted by some models of the early universe and compact astronomical objects
(references are summarized in our previous paper~\cite{Nishizawa1}).
Although the amount of the cosmic GWB is \textit{indirectly} limited by
not only the helium-4 abundance due to big-bang nucleosynthesis~\cite{Maggiore1},
but also measurements of the cosmic microwave background~\cite{Smith1},
direct search experiments for a GWB at very high frequencies should be significant.

We have developed a more sensitive detector for 100-MHz GWs using laser interferometers.
The detector is a pair of synchronous recycling interferometers,
where the synchronous recycling (or resonant recycling) technique was proposed by Drever in the 1980s~\cite{Drever1}.
In our previous papers~\cite{Nishizawa1,Nishizawa2}, we showed that this interferometer is suitable to detect
a GWB at very high frequency with high signal-to-noise ratio (SNR),
and that the SNR can be improved by cross-correlating the outputs of the two interferometers.
In this letter, we report the first results of the search for a stochastic GWB at 100 MHz with the GW detector.

\begin{figure}
\begin{center}
\includegraphics[width=6cm]{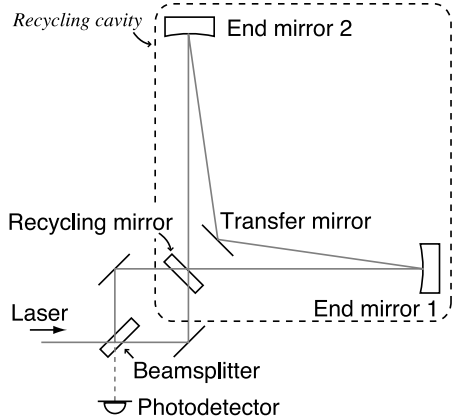}
\caption{Schematic view of a synchronous recycling interferometer.
GW signals are enhanced in the recycling cavity, and detected with the photodetector.}
\label{fig1}
\end{center}
\end{figure}

\textit{Synchronous recycling interferometer.---}
The interferometer has a resonant response to GWs at a specific frequency~\cite{Vinet1,Meers1}.
GW signals are enhanced in a recycling cavity (see Fig.~\ref{fig1}),
which is formed by a recycling mirror (RM), a transfer mirror (TM), and two end mirrors (EM1 and EM2).

The size of the recycling cavity determines the resonant frequency,
where the signal enhancement is proportional to the laser power kept in the cavity.
At the entrance of the interferometer, a laser beam is divided into two orthogonal directions by a beamsplitter (BS).
Thus two beams are incident on the RM, which is a beamsplitter but with relatively high reflectivity.
When the laser frequency is stabilized to the recycling cavity,
the two beams passing through the RM are resonant in the cavity by circulating many times along a common path
in opposite directions (clockwise and counterclockwise).
At the same time, the two circulating beams will experience differential phase shifts due to quadrupole components of GWs.
The phase difference is maximized for the GWs at the same frequency as the free-spectral range $\nu_\mathrm{FSR}$,
the inverse of the round-trip period of the circulating beams.
The phase difference is enhanced as the laser power builds up depending on the finesse of the recycling cavity.
The beams that left the cavity are recombined at the BS so that
the differential components (GW signals) are detected with the photodetector (PD).

\begin{figure}
\begin{center}
\includegraphics[width=8cm]{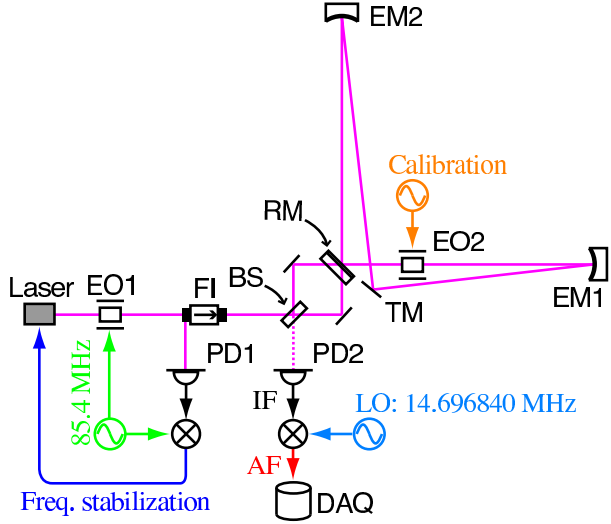}
\caption{(Color online.)
Schematic view of the experimental setup of one of the interferometers.
EO: electro-optic phase modulator; FI: Faraday isolator; DAQ: data acquisition system.
The laser frequency is stabilized to the recycling cavity by the Pound-Drever-Hall technique.
GW signal sidebands are once converted to intermediate-frequency (IF) signals at the PD2.
Then the IF signals are mixed with a local oscillator (LO), and converted to audio-frequency (AF) signals.
The AF signals are recorded with the DAQ.
The EO2 is used to simulate GW signal sidebands for calibration.}
\label{fig2}
\end{center}
\end{figure}

\textit{Experimental setup.---}
We have developed two synchronous recycling interferometers, hereafter called IFO-1 and IFO-2.
For each interferometer (see Fig.~\ref{fig2}), we use a Nd:YAG continuous-wave laser with a wavelength of $1064\,\mathrm{nm}$
and a laser power of $0.5\,\mathrm{W}$.
The laser beam passes through an electro-optic phase modulator (EO1) and a Faraday isolator (FI), and then enters the interferometer.
The recycling cavity is designed to have a baseline length (distance from the RM
to the EM1 or EM2 \footnote{The cavity's optical path is crossed so that it encloses nearly
zero area so as to be insensitive to the Sagnac effect~\cite{Sun1}.}) of $L\simeq 75\,\mathrm{cm}$
so that the GW response is maximized at $\nu_\mathrm{FSR}\equiv c/(4L)\simeq 100\,\mathrm{MHz}$,
where $c$ is the speed of light \footnote{One can construct a narrowband audio-frequency GW detector by adding delay lines or Fabry-Perot cavities in the recycling cavity~\cite{Vinet1,Meers1}.}.
Because this experiment is the first step in the direct detection of a GWB at 100 MHz,
both interferometers are constructed in the air, and each recycling cavity is designed to have a finesse of $\sim 100$;
each RM has relatively low reflectivity (nominal $98.5\%$).
For calibration, we use the EO2 to simulate GWs by modulating the phases of the circulating beams in the cavity.
The size of the Sagnac interferometer, which is formed by the BS, the RM, and two steering mirrors, is relatively small (12.5-cm square optical path),
and thus its GW response is insignificant compared to that of the recycling cavity.

The laser frequency is stabilized to the recycling cavity by the Pound-Drever-Hall technique~\cite{Drever2}.
This technique requires phase-modulation sidebands spaced by a radio frequency (RF)
from the laser-source (carrier) frequency $\nu_0$ in the optical frequency domain.
The RF sidebands at $\nu_0\pm f_\mathrm{RF}$ are induced at the EO1,
where the laser light is phase-modulated at $f_\mathrm{RF}=85.4\,\mathrm{MHz}$.
The PD1 detects the light reflected from the cavity and produces a photocurrent, which contains RF
signals modulated by the relative deviation between the laser and the cavity.
We correct the relative deviation using signals demodulated from the RF signals.

The target GW signals are converted to electrical signals at intermediate frequencies (IF) $\sim 15\,\mathrm{MHz}$ with the PD2,
since it is difficult to make a low-noise photodetector that can respond to signals at very high frequencies ($\sim 100\,\mathrm{MHz}$).
The PD2 produces IF signals at $f_\mathrm{IF}\equiv f_\mathrm{GW}-f_\mathrm{RF}$
in response to the beat between the RF sidebands (also used for the laser stabilization)
and signal sidebands (representation of the GW signals in the optical frequency domain) at $\nu_0\pm f_\mathrm{GW}$,
where $f_\mathrm{GW}$ is the GW frequency.
A small fraction of the RF sidebands leaks to the PD2,
since the splitting ratio of the BS is not exactly balanced in the realistic case.
Otherwise, non-differential components including the RF sidebands are in principle completely reflected into the PD1.
The PD2 is designed to have a band-path filter centered at $f_\mathrm{IF}\sim 15\,\mathrm{MHz}$ with about 1-MHz bandwidth.
Thus the conversion coefficient from GWs to IF signals contains the filter response of the PD2 as well as
the frequency response of the recycling cavity.

\begin{figure}
\begin{center}
\includegraphics[width=8cm]{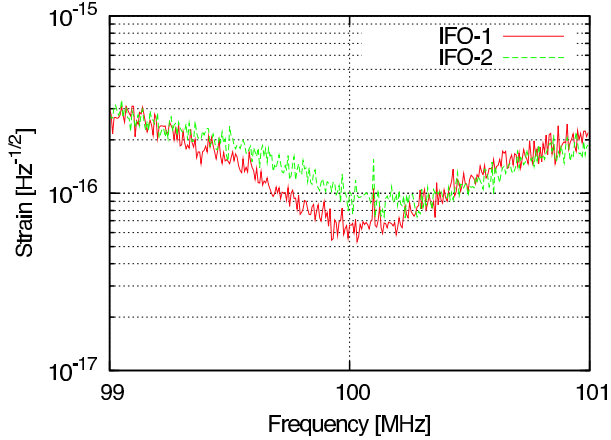}
\caption{(Color online.) Strain sensitivities of the interferometers estimated from the IF signals.
The solid red and dashed green lines represent the sensitivities of IFO-1 and IFO-2, respectively.}
\label{fig3}
\end{center}
\end{figure}

We find the strain sensitivity of each interferometer is about $10^{-16}\,\mathrm{Hz^{-1/2}}$ around $100\,\mathrm{MHz}$ (Fig.~\ref{fig3}) \footnote{There was a concern on environmental electromagnetic noises around $100\,\mathrm{MHz}$.
We use a loop antenna to measure the noise features, and choose a quiet band for the GW search.}.
We estimate the sensitivity from IF signals.
For calibration of the interferometer outputs, we estimate the conversion coefficient from the IF signals $V_\mathrm{IF}$
to the GW amplitudes $h$ by modulating the phases of the circulating beams with the EO2 driven by calibration signals $V_\mathrm{cal}$ \footnote{The calibration signal generator induces noises
as shown in Fig.~\ref{fig3} (the common peaks for both IFOs at about $100.1\,\mathrm{MHz}$ in this case), even when the generator is in a stand-by mode.
We power off the generator during the 1000-second data taking described below.}.
The coefficient is the product of the response of the recycling cavity and the band-path filter of the PD2,
and is estimated by
\begin{equation}
\frac{h(f_\mathrm{GW})}{V_\mathrm{IF}(f_\mathrm{IF})}=\frac{V_\mathrm{cal}(f_\mathrm{GW})}
{V_\mathrm{IF}(f_\mathrm{IF})}A(f_\mathrm{GW})C(f_\mathrm{GW}).
\end{equation}
where $A\,\mathrm{(rad/V)}$ is the measured modulation efficiency of the EO2, and $C\,\mathrm{(strain/rad)}$ is the calculated
conversion coefficient from the phase modulation to the simulated GWs.
The term $C$ is a function of the distance of the EO2 from the RM,
and it is $\sim 25\,\mathrm{cm}$ in our experiment.

The IF signals vary too quickly to be sampled with an inexpensive data acquisition (DAQ) system.
We convert the IF signals to recordable audio-frequency (AF) signals at $f_\mathrm{AF}\equiv f_\mathrm{GW}-(f_\mathrm{c}-\Delta f/2)$
with a local oscillator (LO) at $f_\mathrm{LO}\equiv f_\mathrm{c}-\Delta f/2-f_\mathrm{RF}$,
where we choose
$\Delta f\equiv 6.32\,\mathrm{kHz}$ as a signal bandwidth to be recorded,
and $f_\mathrm{c}\equiv 100.1\,\mathrm{MHz}$ as a center frequency of the bandwidth.
They yield $f_\mathrm{LO}=14.696840\,\mathrm{MHz}$.
For example, GWs at $100.1\,\mathrm{MHz}$ corresponds to AF signals at $3.16\,\mathrm{kHz}$.

\textit{Cross-correlation analysis.---}
Using the outputs of the two interferometers, we have performed a cross-correlation analysis to reduce uncorrelated noises between them
and improve the SNR, the ratio of the GW signals to the interferometer noises.
The analysis method is similar to the method used in LIGO~\cite{Allen1,Abbott3}.
We assume that a GWB is isotropic, unpolarized, stationary, and Gaussian, and it is so small
that the interferometer outputs are dominated by their noises rather than GW signals.
The GWB is often characterized by a normalized energy density spectrum per unit \textit{logarithmic} frequency interval~\cite{Maggiore1}:
$\Omega_\mathrm{gw}(f)\equiv \rho_\mathrm{c}^{-1}\,d\rho(f)/d\ln f,$
where $\rho(f)$ is the cumulative energy density of GWB included below $f\,\mathrm{Hz}$,
and $\rho_\mathrm{c}\equiv 3H_0^2c^2/(8\pi G)$ is the critical energy density of the universe;
here $G$ is the Newton constant, and $H_0\equiv h_{100}\times 100\,\mathrm{km/s/Mpc}$ is the Hubble constant.
In this letter, we also use the form $h_{100}^2\Omega_\mathrm{gw}(f)$,
which is independent of the value of $h_{100}$.

We define a cross-correlation statistic:
\begin{equation}
Z_{12}=\frac{1}{T}\int_{-\infty}^{\infty} \Tilde{x}_1^*(f)\Tilde{x}_2(f)\Tilde{Q}(f)\,df,\label{eqZ}
\end{equation}
where $\Tilde{x}_1$ and $\Tilde{x}_2$ are Fourier components of the signal outputs from IFO-1 and IFO-2, respectively;
$T$ is the observation time period; $\Tilde{Q}$ is the optimal filter that optimizes the SNR of 
an expectation value  (ensemble average) of $Z_{12}$ estimated from available data
(the exact definition of $\Tilde{Q}$ will be given later in Eq.(\ref{eqoptfil})).

The expectation value of $Z_{12}$ and its variance are respectively written as
\begin{align}
&\mu_Z\equiv \left\langle Z_{12} \right\rangle=\frac{3H_0^2}{20\pi^2}
\int_{-\infty}^\infty df \frac{\Omega_\mathrm{gw}(|f|)}{|f|^3}\gamma_{12}(f)\Tilde{Q}(f),\label{eqmuz}\\
&\sigma^2_Z\equiv\left\langle Z_{12}^2 \right\rangle-\left\langle Z_{12} \right\rangle^2
\simeq \frac{1}{4T}\int_{-\infty}^\infty df P_1(|f|)P_2(|f|)|\Tilde{Q}(f)|^2,\label{eqsigma}
\end{align}
where $P_1$ and $P_2$ are the one-sided power spectral densities (PSD) of the noises in IFO-1 and IFO-2, respectively
\footnote{The square root of each PSD is the strain sensitivity.};
and $\gamma_{12}$ is called the \textit{reduced} overlap reduction function.
As the \textit{usual} overlap reduction function in the low-frequency limit~\cite{Christensen1,Flanagan1},
$\gamma_{12}$ represents the reduction of the signal correlation caused by the distance
between the two interferometer sites and the alignment of their arms.
In our experiment, $\gamma_{12}$ $\sim 0.93$ is nearly constant around $100\,\mathrm{MHz}$,
because the two recycling cavities are co-aligned and almost co-located (the distance is $\sim 10\,\mathrm{cm}$)
\footnote{The overlap reduction function, $\gamma(f)$, defined in our previous paper~\cite{Nishizawa2}
is related to $\gamma_{12}$ in the following form:
$\gamma(f)=\gamma_{12}(f)\left[\sin\left(\frac{\pi}{2}\frac{f}{\nu_\mathrm{FSR}}\right)\left/\left(\frac{\pi}{2}\frac{f}{\nu_\mathrm{FSR}}\right)\right.\right]^2$.
Both functions reduce to the usual overlap reduction function in the low-frequency limit: $f\rightarrow 0$.}.

The SNR of the estimation is defined as $\mu_Z/\sqrt{\sigma_Z^2}$.
To optimize the SNR, the filter $\Tilde{Q}$ is chosen as
\begin{equation}
\Tilde{Q}(f)=K\frac{\Omega_\mathrm{gw}(|f|)\gamma_{12}(f)}{|f|^3P_1(|f|)P_2(|f|)},\label{eqoptfil}
\end{equation}
where $K$ is a normalization constant~\cite{Nishizawa2}.
Using this optimal filter, we find that the SNR is written as
\begin{equation}
\textrm{SNR}=\frac{3H_0^2}{10\pi^2}\sqrt{T}\left[\int_{-\infty}^\infty df
\frac{\gamma_{12}^2(f) \Omega^2_\mathrm{gw}(|f|)}{|f|^6P_1(|f|)P_2(|f|)}\right]^{1/2}\label{eqsnr}.
\end{equation}
Thus the SNR in principle increases proportional to $\sqrt{T}$.

The observation time period used for the cross-correlation analysis is 1070.5 seconds.
The data record is divided into $N=439$ segments.
For each segment, the cross-correlation and its uncertainty are calculated based on Eqs.(\ref{eqZ}) and (\ref{eqsigma});
we will refer to the calculated ones as $\hat{Z}_{12}$ and $\hat{\sigma}_Z^2$, respectively.
The ensemble average $\mu_Z$ defined in Eq.(\ref{eqmuz}) is estimated by a weighting average:
\begin{equation}
\hat{\mu}_Z=\hat{\sigma}^2_\mu\sum_{n=1}^{N}\frac{\hat{Z}_{12}^{(n)}}
{\hat{\sigma}_Z^{2\,(n)}},
\end{equation}
where the superscript ``$(n)$'' indicates that the quantity is calculated from the $n$-th segment ($n=1, 2, \dots, N$);
$\sigma_\mu^2$ is the uncertainty of $\hat{\mu}_Z$ and is written as
\begin{equation}
\hat{\sigma}^2_\mu=\left[\sum_{n=1}^N
\frac{1}{\hat{\sigma}_Z^{2\,(n)}}\right]^{-1}.
\end{equation}
As the integration domain in Eq.(\ref{eqZ}), 
we use a range from $2.08\,\mathrm{kHz}$ to $4.19\,\mathrm{kHz}$ for AF signals,
which corresponds to a 2-kHz bandwidth around 100.1 MHz for GW signals.
Because the optimal filter in Eq.(\ref{eqoptfil}) contains $\Omega_\mathrm{gw}(f)$ itself, we need to assume its spectrum in advance.
We assume that the spectrum will be flat in such a narrow bandwidth.

\begin{figure}
\begin{center}
\includegraphics[width=8cm]{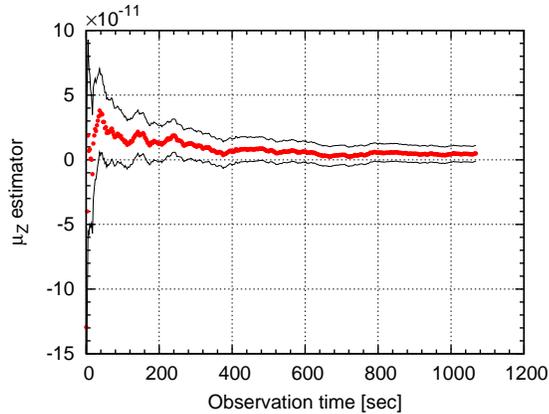}
\caption{(Color online.) Variation of $\hat{\mu}_Z$ with respect to the observation time period is shown as filled red circles.
The area enclosed by black curves is its two-sided $90\%$ confidence interval.}
\label{fig4}
\end{center}
\end{figure}

The variation of $\hat{\mu}_Z$ itself with respect to the observation time period is shown in Fig.\ref{fig4}.
The two curves represent $\hat{\mu}_Z\pm 1.65 \hat{\sigma}_\mu$,
and the area enclosed by the curves is a two-sided $90\%$ confidence interval of $\mu_Z$.
After the 1070.5-second observation, we obtain $\hat{\mu}_Z=4.9\times 10^{-12}$ with $\hat{\sigma}_\mu=3.7\times 10^{-12}$.
The $90\%$ confidence interval $[\hat{\mu}_Z- 1.65 \hat{\sigma}_\mu, \hat{\mu}_Z+1.65 \hat{\sigma}_\mu]$ includes $\mu_Z=0$; in other words, there is a possibility of $\Omega_{\mathrm{gw}}=0$ at 100 MHz.
Instead, we consider an upper limit on the amount of the stochastic GWB at 100 MHz.

We define the upper limit as a one-sided $90\%$ confidence level; in terms of  $\mu_Z$, the upper limit corresponds to $\hat{\mu}_Z+1.28\hat{\sigma}_\mu$.
Then we find $h_{100}^2\Omega_\mathrm{gw}<6\times 10^{25}$ as an upper limit on the stochastic GWB at around 100.1 MHz
from the direct search experiment.
Note that this is also an upper limit on the correlated noises between the two interferometers at this frequency.

\textit{Conclusions.---}
We searched for a stochastic GWB at 100 MHz by laser interferometry.
The GW detector is a pair of synchronous recycling interferometers.
Each interferometer has a strain sensitivity of $\sim 10^{-16}\,\mathrm{Hz^{-1/2}}$ to GWs at $100\,\mathrm{MHz}$.
Using the two interferometers, we directly searched for a stochastic GWB centered at 100.1 MHz with 2-kHz bandwidth in 1070.5 seconds.
We performed a cross-correlation analysis to improve the SNR of the search.
We found $h_{100}^2\Omega_\mathrm{gw}<6\times 10^{25}$ to be an upper limit on the energy density of a stochastic GWB
at 100 MHz.

We plan to improve the GW detector by increasing the finesse of each recycling cavity up to about $4.5\times 10^4$.
For this purpose, each cavity will be constructed in a vacuum with  high-reflectivity mirrors in future. 
Then each interferometer will have a strain sensitivity of about $4.7\times 10^{-21}\,\mathrm{Hz^{-1/2}}$.
For about a one-year observation, we should obtain a tighter upper limit as $h_{100}^2\Omega_\mathrm{gw}\sim 2.8\times 10^{14}$
around $100\,\mathrm{MHz}$ by a cross-correlation analysis with these two interferometers.

This research is supported by Grant-in-Aid for Scientific Research (A) 17204018
from the Ministry of Education, Culture, Sports, Science and Technology.

\newpage 
\bibliography{SrchVHFGWB}

\begin{thebibliography}{14}
\expandafter\ifx\csname natexlab\endcsname\relax\def\natexlab#1{#1}\fi
\expandafter\ifx\csname bibnamefont\endcsname\relax
  \def\bibnamefont#1{#1}\fi
\expandafter\ifx\csname bibfnamefont\endcsname\relax
  \def\bibfnamefont#1{#1}\fi
\expandafter\ifx\csname citenamefont\endcsname\relax
  \def\citenamefont#1{#1}\fi
\expandafter\ifx\csname url\endcsname\relax
  \def\url#1{\texttt{#1}}\fi
\expandafter\ifx\csname urlprefix\endcsname\relax\def\urlprefix{URL }\fi
\providecommand{\bibinfo}[2]{#2}
\providecommand{\eprint}[2][]{\url{#2}}

\bibitem[{\citenamefont{M.~Cruise and Ingley}(2006)}]{Cruise1}
\bibinfo{author}{\bibfnamefont{A.}~\bibnamefont{M.~Cruise}} \bibnamefont{and}
  \bibinfo{author}{\bibfnamefont{R.~M.~J.} \bibnamefont{Ingley}},
  \bibinfo{journal}{Class.\ Quantum.\ Grav.} \textbf{\bibinfo{volume}{23}},
  \bibinfo{pages}{6185} (\bibinfo{year}{2006}).

\bibitem[{\citenamefont{Nishizawa et~al.}(2008)}]{Nishizawa1}
\bibinfo{author}{\bibfnamefont{A.}~\bibnamefont{Nishizawa}}
  \bibnamefont{et~al.}, \bibinfo{journal}{Phys.\ Rev.\ D}
  \textbf{\bibinfo{volume}{77}}, \bibinfo{pages}{022002}
  (\bibinfo{year}{2008}).

\bibitem[{\citenamefont{Maggiore}(2000)}]{Maggiore1}
\bibinfo{author}{\bibfnamefont{M.}~\bibnamefont{Maggiore}},
  \bibinfo{journal}{Phys.\ Rep.} \textbf{\bibinfo{volume}{331}},
  \bibinfo{pages}{283} (\bibinfo{year}{2000}).

\bibitem[{\citenamefont{Smith et~al.}(2006)\citenamefont{Smith, Pierpaoli, and
  Kamionkowski}}]{Smith1}
\bibinfo{author}{\bibfnamefont{T.~L.} \bibnamefont{Smith}},
  \bibinfo{author}{\bibfnamefont{E.}~\bibnamefont{Pierpaoli}},
  \bibnamefont{and}
  \bibinfo{author}{\bibfnamefont{M.}~\bibnamefont{Kamionkowski}},
  \bibinfo{journal}{Phys.\ Rev.\ Lett.} \textbf{\bibinfo{volume}{97}},
  \bibinfo{pages}{021301} (\bibinfo{year}{2006}).

\bibitem[{\citenamefont{W.~P.~Drever}()}]{Drever1}
\bibinfo{author}{\bibfnamefont{R.}~\bibnamefont{W.~P.~Drever}},
  \emph{\bibinfo{title}{Gravitational radiation}}, \bibinfo{note}{edited by N.
  Deruelle and T. Piran (North-Holland, Amsterdam, 1983), pp.321-338}.

\bibitem[{\citenamefont{Nishizawa et~al.}()}]{Nishizawa2}
\bibinfo{author}{\bibfnamefont{A.}~\bibnamefont{Nishizawa}}
  \bibnamefont{et~al.}, \bibinfo{note}{arXiv:0801.4149, ``Optimal Location of
  Two Laser-interferometric Detectors for Gravitational Wave Backgrounds at 100
  MHz''}.

\bibitem[{\citenamefont{Vinet et~al.}(1988)\citenamefont{Vinet, Meers, Man, and
  Brillet}}]{Vinet1}
\bibinfo{author}{\bibfnamefont{J.-Y.} \bibnamefont{Vinet}},
  \bibinfo{author}{\bibfnamefont{B.}~\bibnamefont{Meers}},
  \bibinfo{author}{\bibfnamefont{C.~N.} \bibnamefont{Man}}, \bibnamefont{and}
  \bibinfo{author}{\bibfnamefont{A.}~\bibnamefont{Brillet}},
  \bibinfo{journal}{Phys.\ Rev.\ D} \textbf{\bibinfo{volume}{38}},
  \bibinfo{pages}{433} (\bibinfo{year}{1988}).

\bibitem[{\citenamefont{Meers}(1988)}]{Meers1}
\bibinfo{author}{\bibfnamefont{B.~J.} \bibnamefont{Meers}},
  \bibinfo{journal}{Phys.\ Rev.\ D} \textbf{\bibinfo{volume}{38}},
  \bibinfo{pages}{2317} (\bibinfo{year}{1988}).

\bibitem[{\citenamefont{W.~P.~Drever et~al.}(1983)\citenamefont{W.~P.~Drever,
  L.~Hall, V.~Kowalski, Hough, M.~Ford, and Ward}}]{Drever2}
\bibinfo{author}{\bibfnamefont{R.}~\bibnamefont{W.~P.~Drever}},
  \bibinfo{author}{\bibfnamefont{J.}~\bibnamefont{L.~Hall}},
  \bibinfo{author}{\bibfnamefont{F.}~\bibnamefont{V.~Kowalski}},
  \bibinfo{author}{\bibfnamefont{J.}~\bibnamefont{Hough}},
  \bibinfo{author}{\bibfnamefont{G.}~\bibnamefont{M.~Ford}}, \bibnamefont{and}
  \bibinfo{author}{\bibfnamefont{H.}~\bibnamefont{Ward}},
  \bibinfo{journal}{Appl.\ Phys.\ B} \textbf{\bibinfo{volume}{31}},
  \bibinfo{pages}{97} (\bibinfo{year}{1983}).

\bibitem[{\citenamefont{Allen and D.~Romano}(1999)}]{Allen1}
\bibinfo{author}{\bibfnamefont{B.}~\bibnamefont{Allen}} \bibnamefont{and}
  \bibinfo{author}{\bibfnamefont{J.}~\bibnamefont{D.~Romano}},
  \bibinfo{journal}{Phys.\ Rev.\ D} \textbf{\bibinfo{volume}{59}},
  \bibinfo{pages}{102001} (\bibinfo{year}{1999}).

\bibitem[{\citenamefont{Abbott et~al.}(2007)}]{Abbott3}
\bibinfo{author}{\bibfnamefont{B.}~\bibnamefont{Abbott}} \bibnamefont{et~al.},
  \bibinfo{journal}{Astrophys.\ J.} \textbf{\bibinfo{volume}{659}},
  \bibinfo{pages}{918} (\bibinfo{year}{2007}).

\bibitem[{\citenamefont{Christensen}(1992)}]{Christensen1}
\bibinfo{author}{\bibfnamefont{N.}~\bibnamefont{Christensen}},
  \bibinfo{journal}{Phys.\ Rev.\ D} \textbf{\bibinfo{volume}{46}},
  \bibinfo{pages}{5250} (\bibinfo{year}{1992}).

\bibitem[{\citenamefont{E.~Flanagan}(1993)}]{Flanagan1}
\bibinfo{author}{\bibfnamefont{E.}~\bibnamefont{E.~Flanagan}},
  \bibinfo{journal}{Phys.\ Rev.\ D} \textbf{\bibinfo{volume}{48}},
  \bibinfo{pages}{2389} (\bibinfo{year}{1993}).

\bibitem[{\citenamefont{Sun et~al.}(1996)\citenamefont{Sun, Fejer, Gustafson,
  and Byer}}]{Sun1}
\bibinfo{author}{\bibfnamefont{K.~X.} \bibnamefont{Sun}},
  \bibinfo{author}{\bibfnamefont{M.~M.} \bibnamefont{Fejer}},
  \bibinfo{author}{\bibfnamefont{E.}~\bibnamefont{Gustafson}},
  \bibnamefont{and} \bibinfo{author}{\bibfnamefont{R.~L.} \bibnamefont{Byer}},
  \bibinfo{journal}{Phys.\ Rev.\ Lett.} \textbf{\bibinfo{volume}{76}},
  \bibinfo{pages}{3053} (\bibinfo{year}{1996}).

\end{thebibliography}

\end{document}